\def\simge{{\ \lower2pt\hbox{$\sim$ }\mkern-12mu \raise2pt 
\hbox{$>$}\ }}  

\def\simle{{\ \lower2pt\hbox{$\sim$ }\mkern-18mu \raise2pt 
\hbox{$<$}\ }}  

\def\sumind#1{ {\lower7pt\hbox{$_{#1}$}} \kern-7pt {\hbox{\raise2.5pt
\hbox{$\sum$}}} }

\def\bold#1{\setbox0=\hbox{$#1$}%
      \kern-.02em\copy0\kern-\wd0
      \kern.04em\copy0\kern-\wd0
      \kern-.02em\raise.0433em\box0 }

\def\bdsmall#1{\setbox0=\hbox{$#1$}%
      \kern-.015em\copy0\kern-\wd0
      \kern.03em\copy0\kern-\wd0
      \kern-.015em\raise.0233em\box0 }

\def\ket#1{\vert \, #1 \rangle}

\def\bra#1{\langle #1 \, \vert}

\documentstyle[12pt]{article}
\begin{document}
\rm
\centerline{\bf{ANGULAR DISTRIBUTIONS FOR}}
\centerline{\bf{KNOCKOUT AND SCATTERING OF PROTONS}}  
\centerline{\bf{IN THE EIKONAL APPROXIMATION}}  
     
\vskip1.2truecm
\centerline{\small{A.~ Bianconi}}

\centerline{\small \it{ Dipartimento di Chimica e Fisica per i Materiali, 
Universit\`a di Brescia}}
\centerline{\small \it{ v. Valotti 9, 25133 Brescia, Italy}}

\vskip .5truecm
\centerline{\small{M.~ Radici}}

\centerline{\small \it{ Istituto Nazionale di Fisica Nucleare, 
Sezione di Pavia,}}
\centerline{\small \it{ v. Bassi 6, 27100 Pavia, Italy}}

\vskip2.truecm

\begin{abstract}

The advent of new electron accelerators with few-GeV beam energies 
makes the (e,e$'$p) reaction a promising tool for investigating 
new aspects of the electromagnetic interaction. To this purpose it 
is crucial to set the scale of Final-State Interactions (FSI) at 
high ejectile energies. Usually, the problem is faced by mutuating 
well-established results of the Glauber method in the framework of 
elastic (p,p) scattering. Since the generalization of this 
eikonal approximation to the (e,e$'$p) case is not 
straightforward, we have analyzed the constraints which make the 
comparison a meaningful one, using the 
$^{12}$C(e,e$'$p)$^{11}$B$_{\mathrm{s}1/2}$ and  
$^{11}$B$_{\mathrm{s}1/2}$(p,p) reactions with outgoing-proton 
momenta of 4 GeV/c as a test case. The FSI dominance at large 
deflection angles produces in the distributions a universal 
behaviour resembling the coherent diffractive scattering between 
the ejected proton and the (residual) nucleus. Because of the 
selected sensitivity of the (e,e$'$p) distribution to 
different theoretical ingredients depending on different values 
of the deflection angle (or transverse missing momentum), it is 
argued that the previous comparison with elastic proton 
scattering may represent a convenient tool to disentangle 
effects due to the (hard) electromagnetic vertex from (exotic) 
effects related to the propagation of the struck hadron through 
the nuclear medium.

\end{abstract}

\vskip 1.5cm

\section{Introduction}

With the advent of new electron accelerators, whose beam energy 
will range from the few GeV of CEBAF to the 30 GeV of the planned 
ELFE setup~\cite{elfe}, experiments with electromagnetic probes are expected 
to reveal new physics, particularly on processes like (e,e$'$p) 
scattering~\cite{CEBAF}. Large missing momenta of the residual nucleus will be 
available, where the details of long-range correlations (due to 
the coupling between the motion of the emitten proton and 
collective surface modes of the residual) and of short-range 
correlations (due to the strong nucleon-nucleon interaction) are 
expected to show up in the low and high missing-energy spectrum 
of the residual, respectively~\cite{NIKHEFPb}-\cite{MD94}. 
In addition, because of the high 
momentum and energy transferred to the target, new and unexplored 
features of the electromagnetic hard interaction should appear, 
which are related, for example, to a proper treatment of 
relativistic dynamics and off-shellness~\cite{vanord}. Finally, the subsequent 
propagation of the hadron inside the nuclear medium, usually 
denoted as Final-State Interactions (FSI), is also a central 
ingredient of models aiming to describe exotic effects like 
color transparency~\cite{CT}, if any. 

However, while the paucity of data still prevents from putting 
stringent constraints on the various models for dynamical 
correlations and/or reaction mechanisms at the interaction 
vertex~\cite{NIKHEFPb,MainzO}, the problem of FSI at high projectile energy 
is usually faced by mutuating well-established results obtained in the 
framework of elastic proton scattering. In fact, the Glauber 
approximation~\cite{Glauber} has been extensively used in the past years in the 
analysis of data for (p,p) scattering on complex nuclei~\cite{pp,pprev}. 

But the generalization to the (e,e$'$p) scattering is not 
straightforward, mainly because the kinematics and the state of 
the initial proton are completely different. 
Moreover, the validity of this eikonal approximation, based on a 
completely nonrelativistic formalism, arises from nontrivial 
cancellations among the leading corrections to the lowest-order 
theory of elastic scattering~\cite{wall} and cannot be simply generalized to 
the inelastic case. 

Therefore, after a short review on the general formalism in the 
framework of the Distorted-Wave Impulse Approximation (DWIA) 
(Section II), the constraints which make the comparison between 
FSI in (e,e$'$p) and (p,p) scattering possible, are addressed in 
Section III. 
Firstly, the differences between the two reactions and the choice 
of the proper form of the optical potential 
for distorting the outgoing-proton wave function are discussed. 
Secondly, the restrictions on the 
kinematics and the approximations required to produce similar 
angular distributions are analyzed. Finally, the 
selected sensitivity of the results for (e,e$'$p) scattering to 
different choices of potentials both for the bound and the 
scattering states are considered in Section IV. It is shown that 
at large angles (large values of transverse missing momenta) the 
FSI are the dominant contribution and produce a typical 
diffractive tail very sensitive to the nuclear surface. 

\section{General formalism}

For the scattering of an ultrarelativistic electron with initial 
(final) momentum ${\bold p}_{\mathrm{e}} \  
({\bold p}'_{\mathrm{e}})$, 
while a nucleon is ejected with final momentum ${\bold p}'$, the 
six-fold differential cross section in the one-photon exchange 
approximation reads~\cite{frumou,bgprep} 

\begin{equation}
{ {\mathrm{d}\sigma} \over {\mathrm{d}{\bold p}'_{\mathrm{e}} 
\mathrm{d}{\bold p}'} } = { e^4 \over {8 \pi^2}} {1 \over 
{Q^4 p^{}_{\mathrm{e}} p'_{\mathrm{e}}} } \left( 
\rho_{00} f_{00} + \rho_{11} f_{11} + \rho_{01} f_{01} \cos \alpha 
+ \rho_{1-1} f_{1-1} \cos 2\alpha \right) \  , \label{eq:cross}
\end{equation}
where $Q^2 = {\bold q}^2 - \omega^2$ and ${\bold q} = 
{\bold p}_{\mathrm{e}} - {\bold p}'_{\mathrm{e}}, \  \omega = 
p^{}_{\mathrm{e}} - p'_{\mathrm{e}}$ are the momentum and energy 
transferred to the target nucleus, respectively. The quantities 
$\rho_{\lambda \lambda'}, f_{\lambda \lambda'}$ are expressed 
on the basis of unit vectors 

\begin{eqnarray}
e_0 &= &\left( 1, 0, 0, 0 \right) \nonumber \\
e_{\pm 1} &= &\left( 0, \mp {\textstyle \sqrt{{1 \over 2}}}, - 
{\textstyle \sqrt{{1 \over 2}}} \mathrm{i}, 0 \right) \quad , 
\label{eq:basis}
\end{eqnarray}
which define the longitudinal (0) and transverse $(\pm 1)$ 
components of the nuclear response with respect to the 
polarization of the virtual photon exchanged. The matrix elements 
$\rho_{\lambda \lambda'}$ describe the electrodynamics of the 
leptonic probe, while $f_{\lambda \lambda'}$ depend on $q, 
\omega, p', \cos \gamma = {\bold p}' \cdot {\bold q} / p'q,$ and 
the dependence on the angle $\alpha$, between the 
$({\bold p}',{\bold q})$ plane and the electron scattering plane, 
is explicitely put into evidence. 

The structure functions $f_{\lambda \lambda'}$ are defined in 
terms of bilinear products of the basic ingredient of the 
calculation, the scattering amplitude~\cite{bgprep}

\begin{equation}
J_{\lambda} ({\bold q}) = \int \mathrm{d} {\bold r} \  
\mathrm{e}^{{\scriptstyle \mathrm{i}} {\bdsmall {\scriptstyle q}} 
\cdot {\bdsmall {\scriptstyle r}}} \bra{\Psi_{\mathrm{f}}} 
{\hat J}_{\mu} \cdot e^{\mu}_{\lambda} \ket{\Psi_{\mathrm{i}}} 
\quad ,\label{eq:scattampl}
\end{equation} 
which involves the matrix element of the nuclear charge-current 
density operator ${\hat J}_{\mu}$ between the initial, 
$\ket{\Psi_{\mathrm{i}}}$, and the final, 
$\ket{\Psi_{\mathrm{f}}}$, nuclear states. A natural choice for 
$\ket{\Psi_{\mathrm{f}}}$ is suggested by the experimental 
conditions of the reaction selecting a final state which behaves 
asymptotically as a knocked out nucleon and a residual nucleus in 
a well defined state with energy $E$ and quantum numbers $a$. By 
making the same assumption for the initial state, the two 
specific channels can be projected out of the entire Hilbert 
space by applying a suitable projection operator~\cite{bgprep} to 
$\ket{\Psi_{\mathrm{i}}}$ and $\ket{\Psi_{\mathrm{f}}}$. As a 
result of space truncation, the scattering amplitude is expressed 
in a one-body representation in terms of an appropriate effective 
(one-body) charge-current density operator 
${\hat J}^{\mathrm{eff}}_{\mu}$~\cite{bgprep}:

\begin{equation}
J_{\lambda} ({\bold q}) = \int \mathrm{d} {\bold r} \mathrm{d} \sigma \  
\mathrm{e}^{{\scriptstyle \mathrm{i}} {\bdsmall {\scriptstyle q}} 
\cdot {\bdsmall {\scriptstyle r}}} 
\chi^{\left( -\right)\, *}_{Ea} ({\bold r}, \sigma) \  
{\hat J}^{\mathrm{eff}}_{\mu} \cdot e^{\mu}_{\lambda} \  
\phi_{Ea} ({\bold r}, \sigma) \left[ S_a (E) \right]^{1/2} 
\quad . \label{eq:scattampl1}
\end{equation}
Here $S_a (E)$ is the spectral strength associated with the 
removal process at the excitation energy $E$ of the residual 
nucleus; $\phi_{Ea}$ is eigenfunction of an energy-dependent 
Feshbach optical potential referred to the residual at the energy 
$E$; $\chi^{\left( -\right)}_{Ea}$ is eigenfunction of the optical 
potential at the energy $E+\omega$ and has the boundary conditions 
of incoming wave. The use of an effective current operator in 
eq. (\ref{eq:scattampl1}) takes into account effects due to 
truncation of the Hilbert space and guarantees the orthogonality 
between $\ket{\Psi_{\mathrm{i}}}$ and $\ket{\Psi_{\mathrm{f}}}$~\cite{bccgp82}. 

However, the orthogonality defect is negligible in the standard 
kinematics for (e,e$'$p) reactions and 
${\hat J}^{\mathrm{eff}}_{\mu}$ is usually replaced by 
${\hat J}_{\mu}$~\cite{bccgp82}, which in turn is approximated by a 
nonrelativistic expansion in powers of the inverse nucleon 
mass by means of a Foldy-Wouthuysen canonical transformation~\cite{bgprep}. 
Thus, uncertainties are introduced which depend on the order 
reached in the nonrelativistic expansion and become more 
important with increasing energy~\cite{vanord,gapieka}. But our 
interest is in the analogies between the phenomenology of FSI in 
(p,p) and (e,e$'$p) scattering. Therefore, we have concentrated 
on the properties of the scattering wave 
$\chi^{\left( -\right)}_{Ea}$ and we have considered the 
simplified picture where we retain just the longitudinal 
component ${\hat J}_0$ in the leading order $o(1)$ of the 
nonrelativistic expansion and we neglect the nucleon form factor. 
Consequently, the cross section becomes proportional to
 
\begin{equation} 
\Big \vert \int \mathrm{d} {\bold r} \mathrm{d} \sigma \  
\mathrm{e}^{{\scriptstyle \mathrm{i}} {\bdsmall {\scriptstyle q}} 
\cdot {\bdsmall {\scriptstyle r}} } \chi^{\left( - \right) \, 
*}_{Ea} ({\bold r}, \sigma) \phi_{Ea} ({\bold r}, \sigma) \Big 
\vert^2 \equiv S^{\mathrm{D}}_{Ea} ({\bold q}) \quad ,
\label{eq:specdist} 
\end{equation} 
which is traditionally identified as the ``distorted'' spectral 
density $S^{\mathrm{D}}_{Ea}$~\cite{bgpf79} at the energy $E$ of 
the residual nucleus with a hole with quantum numbers $a$. 

In the framework of the Distorted-Wave Impulse Approximation 
(DWIA) \cite{frumou,bgprep} the socalled spectroscopic amplitudes 
$\phi_{Ea}, \chi^{\left( - \right)}_{Ea}$ are approximated by 
the solutions of eigenvalue problems with single-particle local 
energy-dependent potentials of the Woods-Saxon type. To take into 
account the nonlocality of the original Feshbach potential, these 
eigenfunctions are multiplied by the appropriate Perey factor~\cite{Perey}. 
As for the hole state, in this paper we have considered the 
potential of Comfort and Karp~\cite{bound} for $^{12}$C with the 
quantum numbers of the s$\textstyle{{1 \over 2}}$ shell. The 
scattering wave function $\chi^{\left( - \right)}$ is solution of 
the Schr\"odinger equation 

\begin{equation}
\left( - {\hbar^2 \over {2 m}} \nabla^2 + V \right) \chi = 
E_{\mathrm{cm}} \chi \quad , \label{eq:schroeq}
\end{equation} 
where $m$ is the reduced mass of the proton in interaction with 
the residual nucleus, $E_{\mathrm{cm}}$ is its kinetic energy in 
the cm system and $V$ contains a local equivalent 
energy-dependent optical potential effectively describing the 
residual interaction. 

Eq. (\ref{eq:schroeq}) can be solved for each partial wave of 
$\chi^{\left( - \right)}$ up to a maximum angular momentum 
$L_{\mathrm{max}} (p')$, which satisfies a convergency criterion. 
The boundary condition is such that each incoming 
partial wave coincides asymptotically with the corresponding 
component of the plane wave associated to the proton momentum 
${\bold p}'$. Typically, this method (from now on method A) has 
been applied to (e,e$'$p) scattering with proton momenta 
below 0.5 GeV/c and $L_{\mathrm{max}} < 50$ for a large variety 
of complex optical potentials, including also spin degrees of 
freedom~\cite{bgprep}.  

At higher energies the Glauber method~\cite{Glauber} suggests an 
alternative way (from now on method B) of solving eq. 
(\ref{eq:schroeq}) by linearizing it along the propagation axis 
$\hat z$:

\begin{eqnarray}
{\bold r} &\equiv& z {\displaystyle {{\bold p}' \over p'}} + 
{\bold b} \label{eq:zb} \\
\nabla^2 &\simeq& {\displaystyle {\partial^2 \over 
{\partial z^2}}} \label{eq:nabla} \\
\left( {\partial^2 \over {\partial z^2}} + p'^2 \right) &=& 
{\displaystyle \left( {\partial \over {\partial z}} + \mathrm{i} 
p' \right) \cdot \left( {\partial \over {\partial z}} - \mathrm{i}
p' \right)} \nonumber \\
&\simeq& {\displaystyle 2 \mathrm{i} p' \cdot \left( {\partial 
\over {\partial z}} - \mathrm{i} p' \right)} \quad , 
\label{eq:glau}
\end{eqnarray} 
where ${\bold b}$ describes the degrees of freedom transverse to 
the motion of the struck particle with momentum ${\bold p}'$. 
With this approximation eq. (\ref{eq:schroeq}) becomes 

\begin{equation}
\left( {\partial \over{\partial z}} - \mathrm{i} p' \right) \chi = 
{1 \over {2 \mathrm{i} p'}} \, V \chi \quad . \label{eq:schroglau} 
\end{equation} 
The boundary condition is of incoming unitary flux of plane waves.

\section{Comparison between (e,e$'$p) and (p,p) scattering}

Both methods A and B solve the Schr\"odinger equation for the 
nucleon scattering wave. However, they have been traditionally applied 
to very different reactions, the quasielastic proton knockout and
 the proton elastic scattering, and different energy ranges. Before 
addressing the main goal of this paper, i.e. to test the reliability 
of the Glauber method in (e,e$'$p) reactions and to deduce information on 
FSI by comparison with (p,p) scattering, it is 
necessary to recall these differences and to point out the 
conditions required to allow for a meaningful comparison.     

\subsection{Differences}

Solving the Schr\"odinger equation (\ref{eq:schroeq}) for the 
scattering state implies that the dynamics is calculated in a 
nonrelativistic formalism. Relativistic effects are correctly taken 
into account only in a proper calculation of the kinematics. 
In the case of the application of the Glauber approach to 
unpolarized proton-nucleus elastic scattering, this approximation 
does not seem to produce relevant consequences~\cite{pp,pprev}, 
even if the energies involved would require {\it a priori} a fully 
relativistic treatment. This fact originates from a non trivial 
cancellation among higher-order corrections to the lowest-order 
theory~\cite{wall} and from the observation that the relevant dynamics takes 
place in the transverse plane with respect to the propagation 
axis ${\hat z}$. On the contrary, a fully relativistic description,  
for example of both the bound state and the electromagnetic 
vertex~\cite{amapieka,vanord,gapieka,rost}, seems to play a 
significant role in (e,e$'$p) processes  and the different kinematical 
conditions do not allow for a straightforward generalization of the 
previous results. 

In fact, in (p,p) reactions the angular distribution of the scattered 
proton is caused by ``soft'' diffractive proton-nucleon interactions, 
assuming that rare hard collisions at very large angles are negligible. 
In (e,e$'$p), on the contrary, angular distributions of the 
emitted-proton momentum ${\bold p}'$ with respect to the 
direction of the momentum transfer ${\bold q}$ are possible  
even in a complete absence of proton-nucleon residual interactions, 
because of the Fermi motion of the struck proton when considered in 
its initial bound state. 

Moreover, at increasing energies the physical picture implemented by 
the Glauber method describes a series of ``soft'' rescatterings 
between the target nucleons and the projectile, which is 
approximately considered on-shell. Elastic scattering can be due to 
diffractive regeneration of the on-shell projectile flux. Also 
inelastic intermediate states can play a role, but still they are considered 
on-shell~\cite{gribov}. In the case of proton knockout, the energy 
and momentum transferred to the target can become very high and 
the electromagnetic hard vertex can produce a hadron whose nature 
is quite different from the one of a physical proton. For example, in 
models of color transparency~\cite{hadformlen} the possibility is open for the 
hard production of a hadronic object whose formation length is bigger 
than the nuclear size: this ejectile is simply unable to further 
interact during its propagation through the nuclear medium and 
transforms into an on-shell proton well outside of the nuclear surface. 
Also from the phenomenology of inclusive electron scattering the 
suggestion is put forward that intermediate states with small-mass 
off-shell nucleons are produced by the electromagnetic interaction~\cite{ciosi}. 

To describe this ``exotic'' behaviour of the ejectile it is necessary 
to keep under control the details of its whole scattering wave 
function. Despite of the ambiguities in the optical potential $V$ at 
small $r$ (related to the limits of models for the nucleon-nucleon 
interaction at very short distances), the whole spatial range of 
$\chi^{\left( - \right)} ({\bold r}, \sigma)$ enters the scattering 
amplitude of eq. (\ref{eq:scattampl1}), or alternatively the distorted 
spectral density of eq. (\ref{eq:specdist}). Instead, specific 
assumptions in the Glauber approach allow for the calculation of 
the angular distribution for elastically scattered protons without 
the need of knowing all the details of the projectile wave function 
$\Psi ({\bold r}) \equiv \Psi (r, \theta)$~\cite{Glauber}. Experimental results 
give information on the asymptotic angular distribution of the 
scattered-proton flux with respect to the incoming one, i.e. give 
experimental check only for theoretical calculations of the ratio 
$\vert \Psi (r \rightarrow \infty, \theta) / \Psi (r \rightarrow 
\infty, 180^{\mathrm{o}}) \vert^2$. 

Nevertheless, the comparison with experimental (p,p) angular 
distributions has been quite successful for a large selection of 
target nuclei~\cite{pprev}. While in the case of the only available 
data for (e,e$'$p) at high energies, taken by the NE18 
collaboration~\cite{NE18}, the application of the Glauber model 
in its most straightforward form leads to an overestimation of the 
damping of the outgoing-proton flux at small angles. Several 
interpretations have been proposed to account for this 
discrepancy~\cite{PP92}-\cite{FraStri95}. 
Here, we would like to focus on the features of the distorting 
potential $V(r)$. Since the Glauber approach itself is equivalent 
to the eikonal approximation of method B only for a certain class 
of potentials, a preliminary requirement for any meaningful 
comparison is the proper choice of $V(r)$ in eq. (\ref{eq:schroeq}) 
and eq. (\ref{eq:schroglau}).

\subsection{Choice of the distorting potential}

In the Glauber model $V(r)$ is determined in a parameter-free way 
starting from the elementary free proton-nucleon scattering amplitudes 
at the considered energy~\cite{Glauber}. In DWIA calculations of (e,e$'$p) in 
quasielastic conditions, it has usually a Woods-Saxon form whose 
parameters are fixed by fitting the phase-shifts and the analyzing power 
of elastic (inelastic) (p,p) scattering on the corresponding residual 
nucleus~\cite{bound}. 

In order to set up a potential which can be equivalently used 
with methods A and B, the energy range available to the final proton has 
to be selected. The reliability of the eikonal approximation is 
supposed to increase with increasing ejectile energy~\cite{Glauber}, 
ideally in the limit where $\chi^{\left( - \right)}$ is expanded 
on an infinite number of partial waves. On the other hand, method A 
can be considered reliable only for nucleon energies such that the 
condition $L_{\mathrm{max}} \gg R_{\mathrm{target}} \, p'$ is 
fulfilled, with $R_{\mathrm{target}}$ the radius of the target 
nucleus. Therefore, we have selected outgoing-proton momenta in the 
intermediate range $1 \leq p' \leq 4$ GeV/c and we have solved eq. 
(\ref{eq:schroeq}) up to $L_{\mathrm{max}} = 120$, which matches the 
convergency criterion required. $V(r)$ has the simple Woods-Saxon 
form

\begin{eqnarray}
V(r) &=& \left( U + \hbox{\rm i} W \right) \, {\displaystyle {1 
\over {1 + 
\hbox{\rm e}^{{{r - R} \over a}}}}} \nonumber \\
&\equiv& \left( U + \hbox{\rm i} W \right) \, \rho (r) \quad ,
\label{eq:opt}
\end{eqnarray}
with the parameters adjusted for the $^{12}$C nucleus, i.e. 
$R = 1.2 \times A^{1/3}$ fm and $a = 0.5$ fm. 
The nuclear density $\rho (r)$ defined in eq. (\ref{eq:opt}) is 
normalized such that $\rho (0) = 1$. No spin-orbit contribution 
is taken into account because of the knockout from the s shell of 
$^{12}$C.

At the nucleon momenta here considered, the elementary 
proton-nucleon scattering amplitude is dominated by inelastic 
processes and $V(r)$ is supposed to be mostly sensitive to the 
imaginary well depth $W$~\cite{lech}. However, no phenomenological 
phase-shift analysis is available beyond the inelastic threshold, 
which could constraint $U$ and $W$. In a previous paper~\cite{noiPRC} 
we showed that the $S^{\mathrm{D}}_{\mathrm{s}1/2}$ of eq. 
(\ref{eq:specdist}) for the $^{12}$C(e,e$'$p) reaction at 
$p'=q=1.4$ GeV/c and in perpendicular kinematics (i.e. for $\gamma \neq 0$) 
shows a rather clear insensitivity to the 
sign and magnitude of $U$ for different test choices of $(U,W)$, 
but for huge values $U \gg W$ which are forbidden by the mainly 
absorbitive character of the proton-nucleon amplitude at these 
kinematics. Our conclusion was, therefore, that for $p' \geq 1$ 
GeV/c and confining to perpendicular kinematics one could safely use $U=0$. 
Our choice is not in 
contradiction with the Glauber model, where the ratio $U/W$ 
should equal the ratio between the real and the imaginary parts 
of the average proton-nucleon forward-scattering amplitude, which 
is expected to be small above the inelastic threshold~\cite{lech}.

As it is suggested by eq. (\ref{eq:schroglau}), the Glauber 
approach predicts $W \propto p'$ as far as the proton-nucleon 
total cross section (and, consequently, the damping of the 
proton flux) can be considered constant for different choices 
of $p' \simeq q$, i.e. for small angles. We checked~\cite{noiPRC} 
that the same property holds, with a good approximation, also for 
method A, even below the inelastic threshold. However, in order 
to reproduce the NE18 data, 
a smaller proportionality factor $W/p'$ seems to be required 
with respect to the one indicated by the Glauber model. Various 
interpretations have been suggested to explain this 
discrepancy~\cite{PP92}-\cite{FraStri95}, whose discussion is beyond the 
scope of this paper. Here, we adopt the choice $W \propto p'$ with 
a proportionality factor such as to reproduce the NE18 data, i.e. 
$W = 50 \  p'/1400$ MeV. This choice is equivalent to retaining 
the full Glauber method, but assuming a smaller proton-nucleon 
cross section in nuclear matter than in free space.

\subsection{Analogies}

For the $^{12}$C(e,e$'$p)$^{11}$B$_{\mathrm{s}1/2}$ reaction at 
proton momenta in the range $1 \leq p' \leq 4$ GeV/c we already 
checked~\cite{noiPRC,noiPL} that both methods A and B give 
quite similar angular distribution for 
$S^{\mathrm{D}}_{\mathrm{s}1/2}$. Particularly at $p'=4$ 
GeV/c~\cite{noiPRC} the agreement is impressive and suggests that 
the eikonal approximation of eqs. (\ref{eq:zb})--(\ref{eq:glau}) is 
reliable at these energies. A common feature of both methods is 
that at small angles $\gamma$ (which correspond 
to missing momenta ${\bold p}_{\mathrm{m}} \equiv {\bold p}' - 
{\bold q} \simle p^{}_{\mathrm{Fermi}}$, with 
$p^{}_{\mathrm{Fermi}}$ the Fermi momentum of the target nucleus) 
the distribution is qualitatively dominated 
by the contribution when no FSI are taken into account, i.e. 
in the socalled Plane-Wave Impulse Approximation (PWIA). With a 
good approximation the total result reproduces the 
single-particle momentum distribution of 
the struck proton when in its bound state and an additional 
constant damping. After that threshold, usually around the first 
diffractive minimum of the distribution, the situation changes 
completely. By schematically rewriting eq. (\ref{eq:specdist}) as 

\begin{eqnarray}
S^{\mathrm{D}}_{Ea} ({\bold q}) &\sim& \vert \hbox{\tt PWIA} + 
\hbox{\tt FSI} \vert^2 \nonumber \\
 &=& \vert \hbox{\tt PWIA} \vert^2 + \vert \hbox{\tt FSI} \vert^2 + 2 
\hbox{\rm Re} ( \hbox{\tt PWIA} \cdot \hbox{\tt FSI}^* ) \quad , 
\label{eq:fsi}
\end{eqnarray}
the qualitative picture emerges where for $p_{\mathrm{m}} \sim 
p^{}_{\mathrm{Fermi}}$ the results start becoming sensitive to the 
interference between {\tt PWIA} and {\tt FSI} and for large angles 
$(p_{\mathrm{m}} \gg p^{}_{\mathrm{Fermi}})$ the $\vert 
\hbox{\tt FSI} \vert^2$ contribution dominates producing an 
oscillating diffractive 
pattern which is completely different from the one showed in PWIA 
(see fig. 3 of ref.~\cite{noiPL}).
 In other words, for very large values of transverse 
${\bold p}_{\mathrm{m}}$ the process can be factorized into the 
virtual-photon absorption on a free proton and the subsequent 
coherent diffractive scattering of the struck proton with the 
residual nucleus. Since the diffractive pattern at large angles 
is reminiscent of a similar trend in the proton-nucleus elastic 
scattering~\cite{pp}, it is quite natural to select this kind of 
kinematics and to try to deduce information on FSI by comparison 
between the two different reactions. 

An expression for (p,p) scattering similar to the distorted 
spectral density of eq. (\ref{eq:specdist}) can be written as 

\begin{eqnarray} 
S^{\mathrm{DD}} ({\bold q}) &\equiv &\Big \vert \sumind{Ea} \int 
\mathrm{d} {\bold r} \mathrm{d} \sigma \  \chi^*_{\mathrm{f}} 
({\bold r}, \sigma) \phi^*_{Ea} ({\bold r}, \sigma)  \phi_{Ea} 
({\bold r}, \sigma) \chi_{\mathrm{i}} ({\bold r}, \sigma) \Big 
\vert^2 \nonumber \\
&\simeq &\Big \vert \int \mathrm{d} {\bold r} \mathrm{d} \sigma \  
\chi^*_{\mathrm{f}} ({\bold r}, \sigma) \rho ({\bold r}) 
\chi_{\mathrm{i}} ({\bold r}, \sigma) \Big \vert^2 \quad , 
\label{eq:specddist} 
\end{eqnarray} 
where $\chi_{\mathrm{i}}, \chi_{\mathrm{f}}$ are the distorted 
wave functions for the incoming and outgoing proton flux, 
respectively, and the sum runs over all the possible discrete 
states $\phi_{Ea}$ with energy $E$ and quantum numbers $a$, that 
the intermediate proton can form with the target. The 
$\rho ({\bold r})$ is, in principle, the diagonal part of the 
density matrix; in practice, it is approximated by the nuclear 
density of the target. Eq. (\ref{eq:specddist}) is the simplest 
expression that can be conceived to build the cross section for 
(p,p) scattering. Many other corrections have been presented in 
the literature~\cite{pprev}, which would correspond to further 
improvements in the treatment of FSI in eq. (\ref{eq:specdist}), 
and therefore are disregarded. 

Assuming the validity of the eikonal approximation, the density 
$\rho (r)$ becomes proportional to the potential $V(r)$ entering 
eq. (\ref{eq:schroglau}), whose solutions $\chi_{\mathrm{i}}, 
\chi_{\mathrm{f}}$ are 

\begin{eqnarray}
\chi_{\mathrm{i}} ({\bold r}) &= &\mathrm{e}^{{\scriptstyle 
\mathrm{i}} {\bdsmall {\scriptstyle p}}_{\scriptscriptstyle{
\mathrm{i}}} \cdot {\bdsmall {\scriptstyle r}} } \ 
\mathrm{e}^{{\scriptstyle C \int_{\scriptscriptstyle - 
\infty}^{\scriptscriptstyle z} \rho} \left( {\bdsmall 
{\scriptstyle r}}_{\scriptscriptstyle{\perp}}, \, 
z'_{\scriptscriptstyle \mathrm{i}} \right) 
{\scriptstyle \mathrm{d} z'_{\scriptscriptstyle \mathrm{i}}}} 
\nonumber \\
\chi_{\mathrm{f}} ({\bold r}) &= &\mathrm{e}^{{\scriptstyle 
\mathrm{i}} {\bdsmall {\scriptstyle p}}_{\scriptscriptstyle{
\mathrm{f}}} \cdot {\bdsmall {\scriptstyle r}} } \ 
\mathrm{e}^{{\scriptstyle C \int_{\scriptscriptstyle 
z}^{\scriptscriptstyle + \infty} \rho} \left( {\bdsmall 
{\scriptstyle r}}_{\scriptscriptstyle{\perp}}, \, 
z'_{\scriptscriptstyle \mathrm{f}} \right) 
{\scriptstyle \mathrm{d} z'_{\scriptscriptstyle \mathrm{f}}}} 
\quad , \label{eq:chieik}
\end{eqnarray}
where $C$ is a constant factor relating $V(r)$ to $\rho(r)$ and  
${\bold r}_{\perp}$ describes the degrees of freedom in the 
transverse plane with respect to the propagation axis 
${\hat z}'_{\mathrm{i}}, {\hat z}'_{\mathrm{f}}$, which are taken 
parallel to the momenta ${\bold p}_{\mathrm{i}}, 
{\bold p}_{\mathrm{f}}$ of the incoming and outgoing proton, 
respectively. 

Since at high proton momenta the angular deviation from the initial 
trajectory is usually small, the integrals in eq. (\ref{eq:chieik}) 
can be computed in the average direction ${\hat z}' = \left( 
{\hat z}'_{\mathrm{i}} + {\hat z}'_{\mathrm{f}} \right) /2$. Therefore, 
eq. (\ref{eq:specddist}) becomes 
\begin{eqnarray}
S^{\mathrm{DD}} ({\bold q}) &= &\Big \vert \int \mathrm{d} {\bold r} 
\rho (r) \  \mathrm{e}^{{\scriptstyle - \mathrm{i}} \left( 
{\bdsmall {\scriptstyle p}}_{\scriptscriptstyle{\mathrm{f}}} - 
{\bdsmall {\scriptstyle p}}_{\scriptscriptstyle{\mathrm{i}}} 
\right) \cdot {\bdsmall {\scriptstyle r}} } \ 
\mathrm{e}^{{\scriptstyle C \int_{\scriptscriptstyle - 
\infty}^{\scriptscriptstyle + \infty} \rho} \left( 
{\bdsmall {\scriptstyle r}}_{\scriptscriptstyle{\perp}}, \, z' 
\right) {\scriptstyle \mathrm{d} z'}} \Big \vert^2 \nonumber \\
&\equiv &\Big \vert \int \mathrm{d} {\bold r} \rho (r) \  
\mathrm{e}^{{\scriptstyle - \mathrm{i}} {\bdsmall 
{\scriptstyle p}}_{\scriptscriptstyle{\mathrm{m}}}
\cdot {\bdsmall {\scriptstyle r}} } \ \mathrm{e}^{{\scriptstyle C 
\int_{\scriptscriptstyle - \infty}^{\scriptscriptstyle + \infty} 
\rho} \left( {\bdsmall {\scriptstyle r}}_{\scriptscriptstyle{\perp}}, \, z' 
\right) {\scriptstyle \mathrm{d} z'}} \Big \vert^2 \quad , \label{eq:ddisteik}
\end{eqnarray}
which produces the same results of the Glauber standard 
expression~\cite{Glauber}
\begin{equation}
\int \mathrm{d} {\bold b} \  \mathrm{e}^{{\scriptstyle - \mathrm{i}} {\bdsmall 
{\scriptstyle p}}_{\scriptscriptstyle{\mathrm{m}}} \cdot {\bdsmall 
{\scriptstyle b}} } \ \left[ 1 - \mathrm{e}^{{\scriptstyle C 
\int_{\scriptscriptstyle - \infty}^{\scriptscriptstyle + \infty} 
\rho} \left( {\bdsmall {\scriptstyle r}}_{\scriptscriptstyle{\perp}}, \, z' 
\right) {\scriptstyle \mathrm{d} z'}} \right] \quad . \label{eq:eikstand}
\end{equation}
Here, ${\bold p}_{\mathrm{m}} = {\bold p}_{\mathrm{f}} - 
{\bold p}_{\mathrm{i}}$ represents now the difference between 
the final and initial momenta of the proton, respectively, and is 
perpendicular to the average propagation axis ${\hat z}'$. Thus 
confirming the previous qualitative findings, a meaningful comparison 
with the (e,e$'$p) case is possible only for kinematics with 
large values of transverse missing momenta ${\bold p}_{\mathrm{m}} 
= {\bold p}' - {\bold q}$. In fig. 1 the $S^{\mathrm{DD}}$ of 
eq. (\ref{eq:ddisteik}) is shown by the long-dashed curve for 
the $^{11}$B$_{\mathrm{s}1/2}$(p,p) reaction at 
$p^{}_{\mathrm{f}} = 4$ GeV/c. 

By applying the same eikonal approximation to the distorted 
spectral density for (e,e$'$p), eq. (\ref{eq:specdist}) becomes 

\begin{eqnarray}
S^{\mathrm{D}}_{\mathrm{s}1/2} ({\bold q}) &= &\Big \vert \int 
\mathrm{d} {\bold r} \phi_{\mathrm{s}1/2} ({\bold r}) \  
\mathrm{e}^{{\scriptstyle - \mathrm{i}} \left( {\bdsmall 
{\scriptstyle p}}' - {\bdsmall {\scriptstyle q}} \right) \cdot 
{\bdsmall {\scriptstyle r}} } \ \mathrm{e}^{{\scriptstyle C 
\int_{\scriptscriptstyle z}^{\scriptscriptstyle + \infty} 
\rho} \left( {\bdsmall {\scriptstyle r}}_{\scriptscriptstyle{\perp}}, 
\, z' \right) {\scriptstyle \mathrm{d} z'}} \Big \vert^2 \nonumber \\
&\equiv &\Big \vert \int \mathrm{d} {\bold r} \phi_{\mathrm{s}1/2} 
({\bold r})  \  \mathrm{e}^{{\scriptstyle - \mathrm{i}} 
{\bdsmall {\scriptstyle p}}_{\scriptscriptstyle{\mathrm{m}}} \cdot 
{\bdsmall {\scriptstyle r}} } \ \mathrm{e}^{{\scriptstyle C 
\int_{\scriptscriptstyle z}^{\scriptscriptstyle + \infty} 
\rho} \left( {\bdsmall {\scriptstyle r}}_{\scriptscriptstyle{\perp}}, 
\, z' \right) {\scriptstyle \mathrm{d} z'}} \Big \vert^2 \quad . 
\label{eq:disteik}
\end{eqnarray}
The first difference between eq. 
(\ref{eq:disteik}) and eq. (\ref{eq:ddisteik}) is the 
$z$-dependence of the integral involving the optical potential. If 
in eq. (\ref{eq:disteik}) ${\bold p}_{\mathrm{m}}$ is chosen to 
be perpendicular to ${\hat z}$, the Fourier transform will be 
largely unaffected by the $z$-dependence of the integral and 
$S^{\mathrm{D}}_{\mathrm{s}1/2}$ can be approximated by 

\begin{equation}
S^{\mathrm{D}}_{\mathrm{s}1/2} ({\bold q}) \simeq \Big \vert 
\int \mathrm{d} {\bold r} \phi_{\mathrm{s}1/2}
({\bold r})  \  \mathrm{e}^{{\scriptstyle - \mathrm{i}} 
{\bdsmall {\scriptstyle p}}_{\scriptscriptstyle{\mathrm{m}}} \cdot 
{\bdsmall {\scriptstyle r}} } \ \mathrm{e}^{{\scriptstyle {C 
\over 2} \int_{\scriptscriptstyle - \infty}^{\scriptscriptstyle + 
\infty} \rho} \left( {\bdsmall {\scriptstyle 
r}}_{\scriptscriptstyle{\perp}}, \, z' \right) 
{\scriptstyle \mathrm{d} z'}} \Big \vert^2 \quad .
\label{eq:disteikca}
\end{equation}
In fig. 1 the solid and short-dashed curves represent eq. 
(\ref{eq:disteik}) and eq. (\ref{eq:disteikca}) for the 
$^{12}$C(e,e$'$p)$^{11}$B$_{\mathrm{s}1/2}$ reaction at 
$p' = q = 4$ GeV/c, respectively. The similarity of the two curves 
confirms the insensitivity to the longitudinal position of the 
knockout point $z$. It must be stressed that this is justified only 
for ${\bold p}_{\mathrm{m}} \perp {\hat z}$. Assuming that at high 
energies and momenta the $S^{\mathrm{D}}_{\mathrm{s}1/2}$ is less 
sensitive to the details of the bound state $\phi_{\mathrm{s}1/2}$ 
and is dominated by the exponential factors, a very close similarity 
can be recovered between eq. (\ref{eq:ddisteik}) and eq. 
(\ref{eq:disteikca}). The corresponding long-dashed and solid 
curves in fig. 1 show, after the threshold of the first 
diffractive minimun where the $\vert \hbox{\tt FSI} \vert^2$ 
contribution in eq. (\ref{eq:fsi}) starts dominating, the same 
universal angular pattern, thus confirming the previous 
assumption on the FSI dominance at large energies. 

The situation can be summarized as follows. In the absence of 
exotic effects, the FSI for the (e,e$'$p) reaction become dominant 
approximately beyond deflection angles $\gamma$ such that the 
missing momentum ${\bold p}_{\mathrm{m}}$ exceeds the 
$p^{}_{\mathrm{Fermi}}$ of the target nucleus. The angular 
distribution for large values of transverse 
${\bold p}_{\mathrm{m}}$ is completely different from the PWIA 
result and shows the same universal diffractive  
pattern of the distribution of protons elastically scattered by 
the same residual nucleus and at the same energy and momentum. 

Therefore, since no information was put inside the matrix 
elements of eq. (\ref{eq:disteik}) 
about the interaction vertex, any deviation from the previous 
picture has to be ascribed to the details of the (hard) virtual-photon 
absorption in nuclear medium and to the modifications that can induce 
on the struck hadron. For example, it has been argued~\cite{Kope} 
for (e,e$'$p) that, because of inelastic corrections, at 
increasing energy a rise of the nuclear transparency is to be 
expected, that would be hardly distinguishable from effects like 
color transparency. Since this kind of inelastic 
corrections is one of the higher-order ingredients adopted to 
improve the (p,p) elastic cross section of eq. 
(\ref{eq:specddist})~\cite{gribov}, the comparison between the two reactions in 
the kinematics above specified could be of much help. In general, 
the signature of any possible color transparency phenomenon in 
hard (e,e$'$p) scattering is that the nuclear response should 
look more similar to the PWIA result. The traditional strategy 
has been so far to search for variations of the nuclear damping 
in the outgoing-proton flux, particularly at small missing 
momenta~\cite{NE18}. However, very precise and unambiguous results must be 
obtained to this purpose. From previous comments, it could be 
equally convenient to analyze the angular distribution for 
completely exclusive reactions, because FSI can significantly 
``distort'' the PWIA result. Moreover, from the comparison with 
the diffractive tail of the corresponding elastic (p,p) 
distribution further insight into the reaction mechanism of the 
(hard) electromagnetic vertex could be gained.

\section{Properties of FSI for large-angle distributions}

It has already been observed that in the angular distribution for 
(e,e$'$p) scattering a special role is played by the 
$p^{}_{\mathrm{Fermi}}$ of the target nucleus. In fact, for angles 
corresponding to transverse missing momenta larger than 
$p^{}_{\mathrm{Fermi}}$ the relation $\hbox{\tt FSI} \gg 
\hbox{\tt PWIA}$ holds and the shape of the curve is determined by the 
rescatterings of the hit hadron.  

In fig. 2 the $S^{\mathrm{D}}_{\mathrm{s}1/2}$ is shown by 
the dashed line for the 
$^{12}$C(e,e$'$p)$^{11}$B$_{\mathrm{s}1/2}$ reaction at 
$p'=q=4$ GeV/c with the bound state taken from the solution of 
the Woods-Saxon potential of Comfort and Karp~\cite{bound} and 
with the optical potential described in subsection 3.2. The solid 
curves are produced by varying the well radius $R$ of the optical 
potential. The small-angle part of the distribution (around the 
first minimum and following secondary maximum, i.e. for 
$p_{\mathrm{m}} \simle 2 p_{\mathrm{Fermi}}$, which is around 
$\gamma = 6^{\mathrm{o}}$ for $p'=4$ GeV/c) is not very much 
affected, while the large-angle diffractive pattern is 
significantly modified both in the size and in the frequency of 
the secondary maxima. On the contrary, no significant change is 
observed when keeping everything fixed but the imaginary depth 
$W$ in eq. (\ref{eq:opt}), as it is evident from fig. 3. Assuming 
that the residual nucleus can be represented, in a simplified 
picture, as a ``nuclear 
lense'', from fig. 2 it can be deduced that modifying the 
size of the lense changes the diffractive shape of the beam of 
particles scattered at large angles. However, the average slope, 
which can be identified as the tangent to the distribution in 
the secondary maxima, is not modified. Moreover, a regular 
oscillatory pattern is due to interference among the fluxes of 
particles scattered by a discrete (periodical) structure of 
scatterers, typically point-like sources or a lattice. Diffraction 
from a continuous structure would cause a distribution with a 
single central maximum. In the case of (e,e$'$p), the obvious 
identification follows between the structure of scatterers and the 
nucleons inside the residual nucleus. But in the formalism leading 
to eq. (\ref{eq:specdist}) there is no signature of the 
many-body aspect of the residual interaction. The optical model 
is, in fact, a mean-field approximation to the problem of FSI 
with smoothly varying properties.

However, the diffuseness $a$ of the Woods-Saxon well (see eq. 
(\ref{eq:opt})) is the only parameter that introduces into the 
problem a dimensional length of the order of the nucleon size, 
which is in turn very similar to the range of nucleon-nucleon 
correlations. Inspection of fig. 4, where 
$S^{\mathrm{D}}_{\mathrm{s}1/2}$ is calculated in the same 
conditions as in fig. 2, shows that varying $a$ not 
only modifies the size, but also the average slope of the 
angular distribution. The dashed line here corresponds to 
the dashed line in fig. 2. It is evident that large-angle emissions 
are largely affected by the nucleon-nucleon interactions taking 
place in the nuclear surface. Also the short-distance structure 
of the internal nuclear medium is important, but small volumes in 
the nuclear interior can be considered roughly isotropic and 
unable to select a preferred direction (as it is usually assumed in 
the Local Density Approximation). 

By combining the previous observations about the sensitivity of the 
results to $R$ and $a$, we can deduce first that a small spatial region 
of size $a$ on the nuclear surface is responsible for the overall 
feature of the large-angle distribution in momentum space. 
Secondly, in eq. (\ref{eq:ddisteik}) the surface oscillations of 
$\rho (r)$ produce high-frequency components in momentum space, 
among which only those close to ${\bold p}_{\mathrm m}$ are 
emphasized in the Fourier transform. This corresponds to selecting 
only two surface regions of size $a$, which can contribute to the 
emission of a nucleon with initial missing momentum 
${\bold p}_{\mathrm m}$. They can be identified through the intersection 
between the direction $\hat {\bold p}_{\mathrm m}$ and the nuclear 
surface. The diffractive pattern can, therefore, be interpreted 
as the quantum interference between the fluxes emerging from these two 
regions. In fact, from fig. 2 it turns out that the finer details 
of the oscillations in momentum space are sensitive to the nuclear 
size $R$, or equivalently to the relative distance between the 
two regions, which is much bigger than their size $a$. 

From this picture the findings in ref.~\cite{GerIre} are confirmed 
that, by a suitable modification of $R, a$ in the optical 
potential in a way compatible with the constraints dictated by 
phase-shift analysis, most details of final-state rescattering 
in (e,e$'$p) at large angles can be effectively 
reproduced by a mean-field optical model.

Finally, it must be stressed that for all these results it is 
crucial that the residual nucleus be in a well defined state. 
Only in this case its internal structure can be coherently tested 
by the ejectile. Energy-integrated distributions (like 
semi-inclusive (e,e$'$p) reactions~\cite{niko,andrea}) can test by definition 
only the average behaviour of the emitted proton, thus leading to very 
different angular shapes.

\section{Conclusions}

It has been shown elsewhere~\cite{noiPRC,noiPL} that for the 
$^{12}$C(e,e$'$p)$^{11}$B$_{\mathrm{s}1/2}$ reaction at 
proton momenta $1 \leq p' \leq 4$ GeV/c (relevant to the planned 
experiments at CEBAF) the eikonal approximation to the scattering 
wave of the ejectile produces angular distributions very similar 
to the ones obtained when the complete second-order differential 
equation is solved up to 120 partial waves. Assuming this 
approximation as a reliable one, it has been here demonstrated 
that the large-angle part of this distribution and the 
corresponding one for 
$^{11}$B$_{\mathrm{s}1/2}$(p,p)$^{11}$B$_{\mathrm{s}1/2}$ 
elastic scattering have a universal feature corresponding to a 
coherent diffractive scattering of the outgoing proton from the 
$^{11}$B$_{\mathrm{s}1/2}$ excited nucleus. This is due to the 
dominance of FSI in this kinematical region. Because of the 
nontrivial differences between the two reactions, the comparison 
is meaningful only for missing momenta with a large transverse 
component with respect to the momentum transfer ${\bold q}$. 
With these constraints, it is argued that it may represent a 
more convenient tool to disentangle effects due to the (hard) 
electromagnetic vertex from (exotic) effects related to the 
propagation of the struck hadron through the nuclear medium. 
While the small-angle part of the distribution is affected mainly 
by the single-particle momentum distribution of the emitted proton 
when in its bound state (already accessible in the PWIA 
approximation), the large-angle part shows a marked sensitivity 
to the dimensional parameters of the well of the residual 
potential, particularly to its surface thickness which is related 
to the dimensional scale of the short-range nucleon-nucleon 
interaction. 

\vspace{.6cm}

We would like to thank O. Benhar, S. Boffi, S. Jeschonnek, N.N. 
Nikolaev and S. Simula for many stimulating discussions.


\newpage


\centerline{Captions}

\vspace{2cm}

\begin{itemize}

\item[Fig. 1 - ] The solid line represents the distorted spectral 
density $S^{\mathrm{D}}_{\mathrm{s} 1/2}$ in the eikonal 
approximation for the 
$^{12}\mathrm{C(e,e}'\mathrm{p)}^{11}\mathrm{B}_{\mathrm{s} 1/2}$ 
reaction at $p' = q = 4$ GeV/c for various values of transverse 
missing momentum and for a purely imaginary optical potential with 
depth $W = 50 \  p'/p_{\mathrm{o}}$ MeV, with $p_{\mathrm{o}} = 
1.4$ GeV/c. The bound state is derived from the potential of 
Comfort and Karp~\cite{bound}. The short-dashed line shows the 
result when the further approximation of eq. (\ref{eq:disteikca}) 
is applied (see text). The long-dashed line refers to the 
transition probability $S^{\mathrm{DD}}$ of eq. 
(\ref{eq:ddisteik}) (see text) for the 
$^{11}$B$_{\mathrm{s}1/2}$(p,p)$^{11}$B$_{\mathrm{s}1/2}$ 
reaction in the same kinematics. 

\end{itemize}

\vspace{1cm}

\begin{itemize}

\item[Fig. 2 - ]  The distorted spectral density 
$S^{\mathrm{D}}_{\mathrm{s} 1/2}$ for the 
$^{12}\mathrm{C(e,e}'\mathrm{p)}^{11}\mathrm{B}_{\mathrm{s} 1/2}$ 
reaction in the same kinematical conditions as in fig. 1. The dashed line 
corresponds to the solid line in fig. 1. The upper (at $\gamma 
\sim 6^{\mathrm{o}}$) solid lines are obtained when reducing the 
well radius of the optical potential by $17\%$ and $8\%$; the 
lower one when increasing it by $8\%$.  

\end{itemize}

\vspace{1cm}

\begin{itemize}

\item[Fig. 3 - ] The distorted spectral density 
$S^{\mathrm{D}}_{\mathrm{s} 1/2}$ for the 
$^{12}\mathrm{C(e,e}'\mathrm{p)}^{11}\mathrm{B}_{\mathrm{s} 1/2}$ 
reaction in the same kinematical conditions as in fig. 1, but with a variable 
imaginary depth $W$ of the optical 
potential. The dashed line is obtained with $W = 150$ MeV, which 
corresponds approximately to the solid line in fig. 1. The upper 
and lower (at $\gamma = 0^{\mathrm{o}}$) solid lines correspond 
to $W = 100$ and $200$ MeV, respectively.   

\end{itemize}

\vspace{1cm}

\begin{itemize}

\item[Fig. 4 - ] The distorted spectral density 
$S^{\mathrm{D}}_{\mathrm{s} 1/2}$ for the 
$^{12}\mathrm{C(e,e}'\mathrm{p)}^{11}\mathrm{B}_{\mathrm{s} 1/2}$ 
reaction in the same kinematical conditions as in fig. 1. The dashed line 
corresponds to the solid line in fig. 1. The lower (at $\gamma 
\sim 10^{\mathrm{o}}$) solid line is obtained when reducing the 
diffuseness of the optical potential by $50\%$; the upper one 
when increasing it by $50\%$.

\end{itemize}

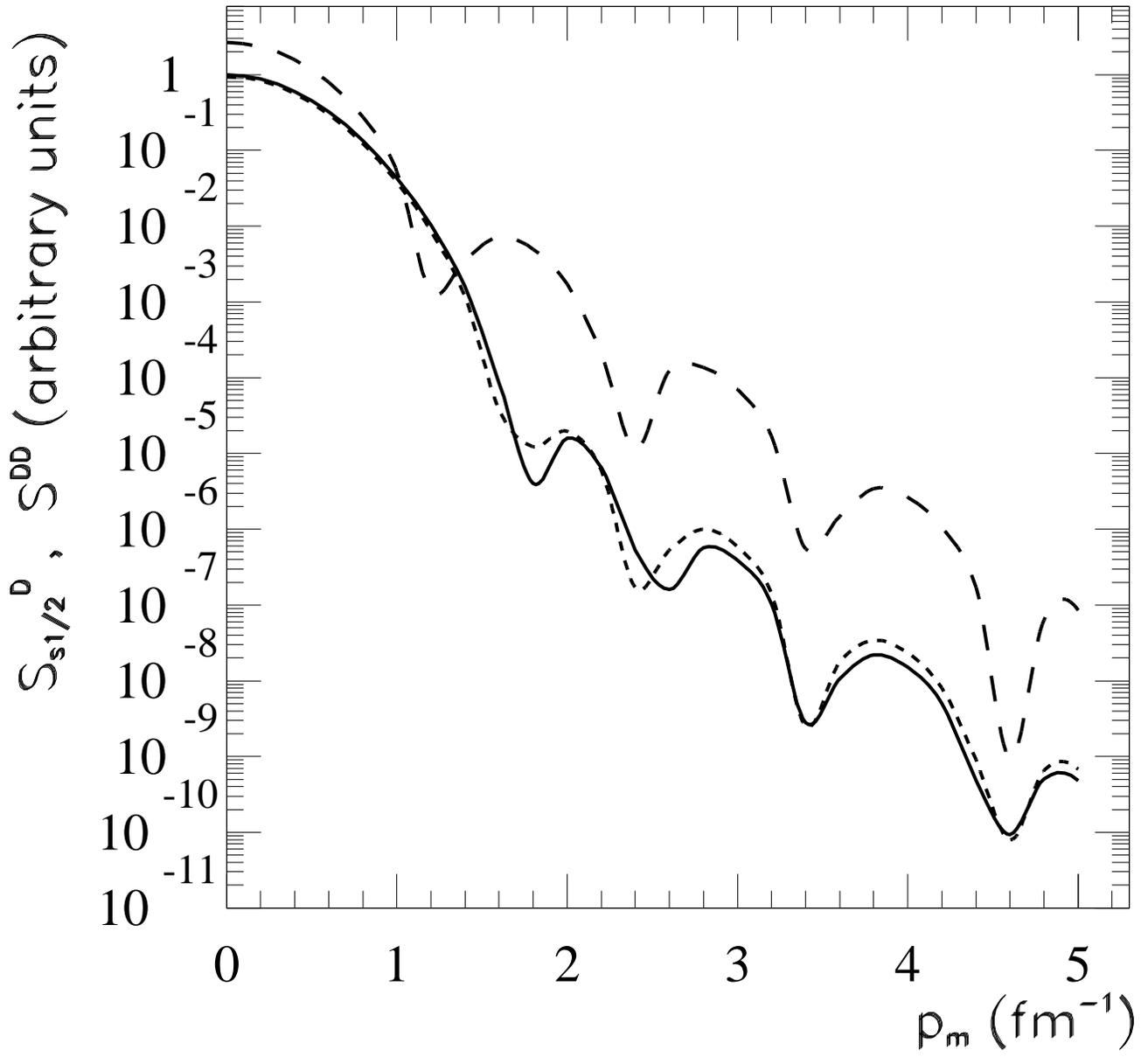
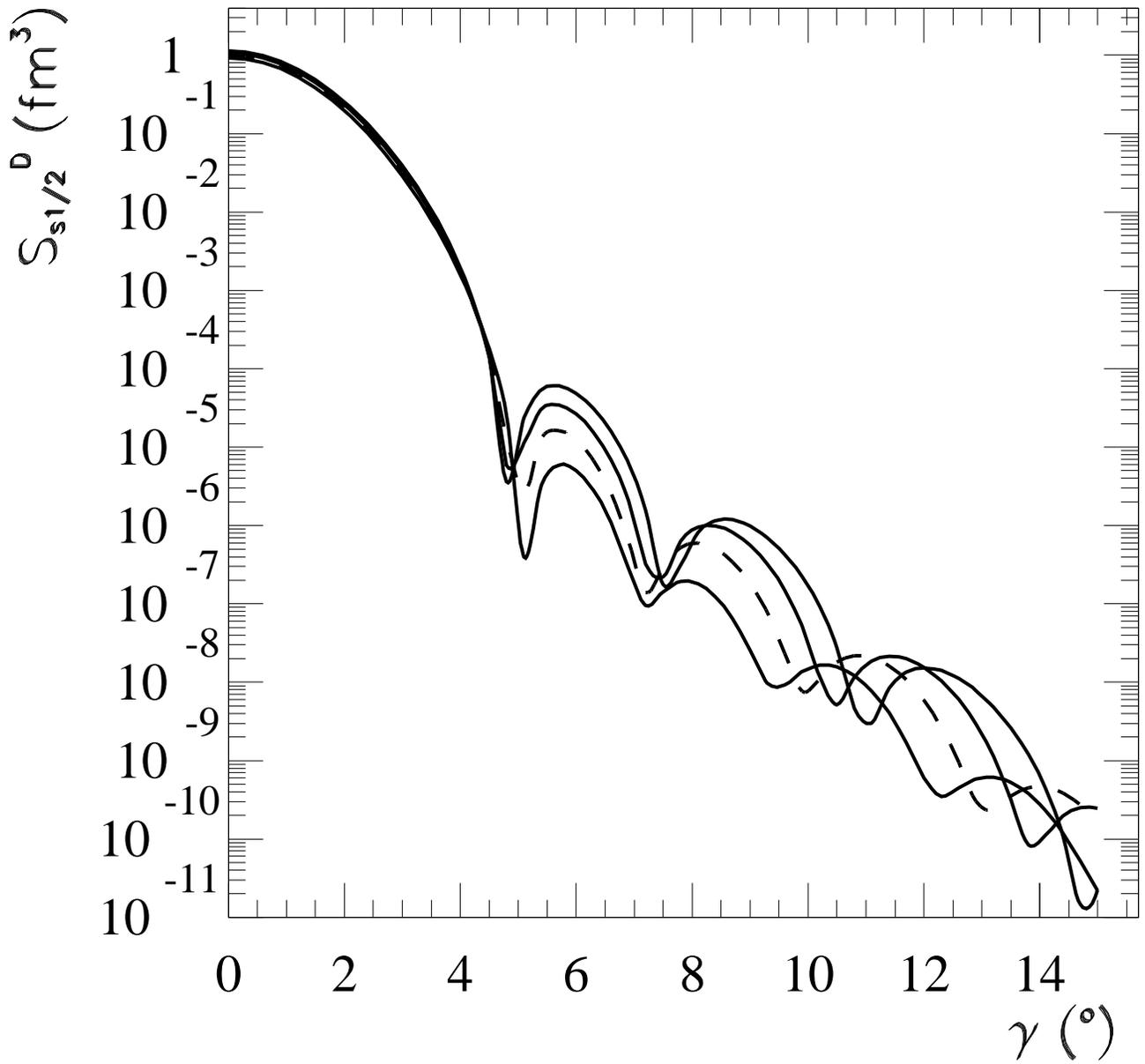
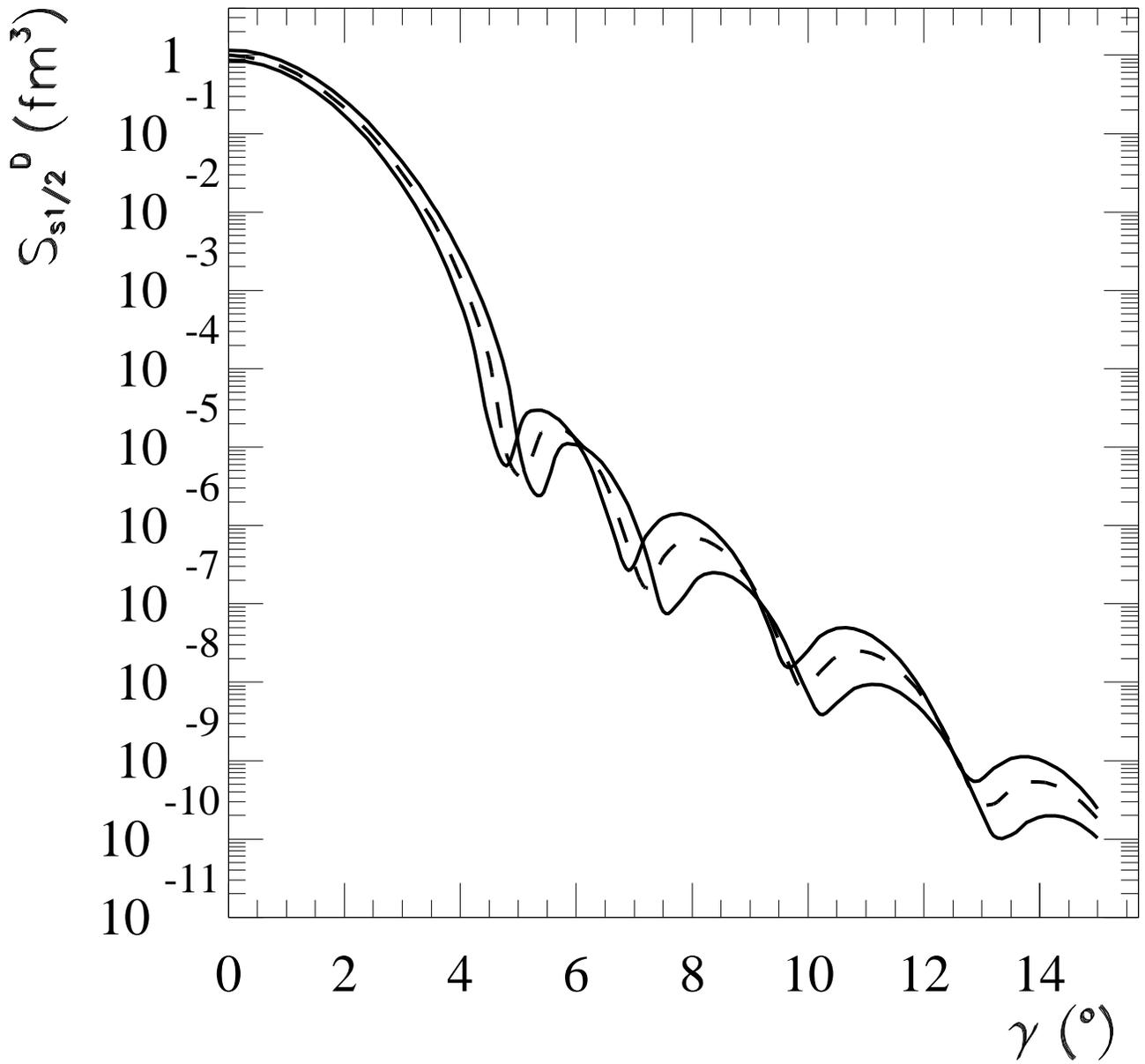
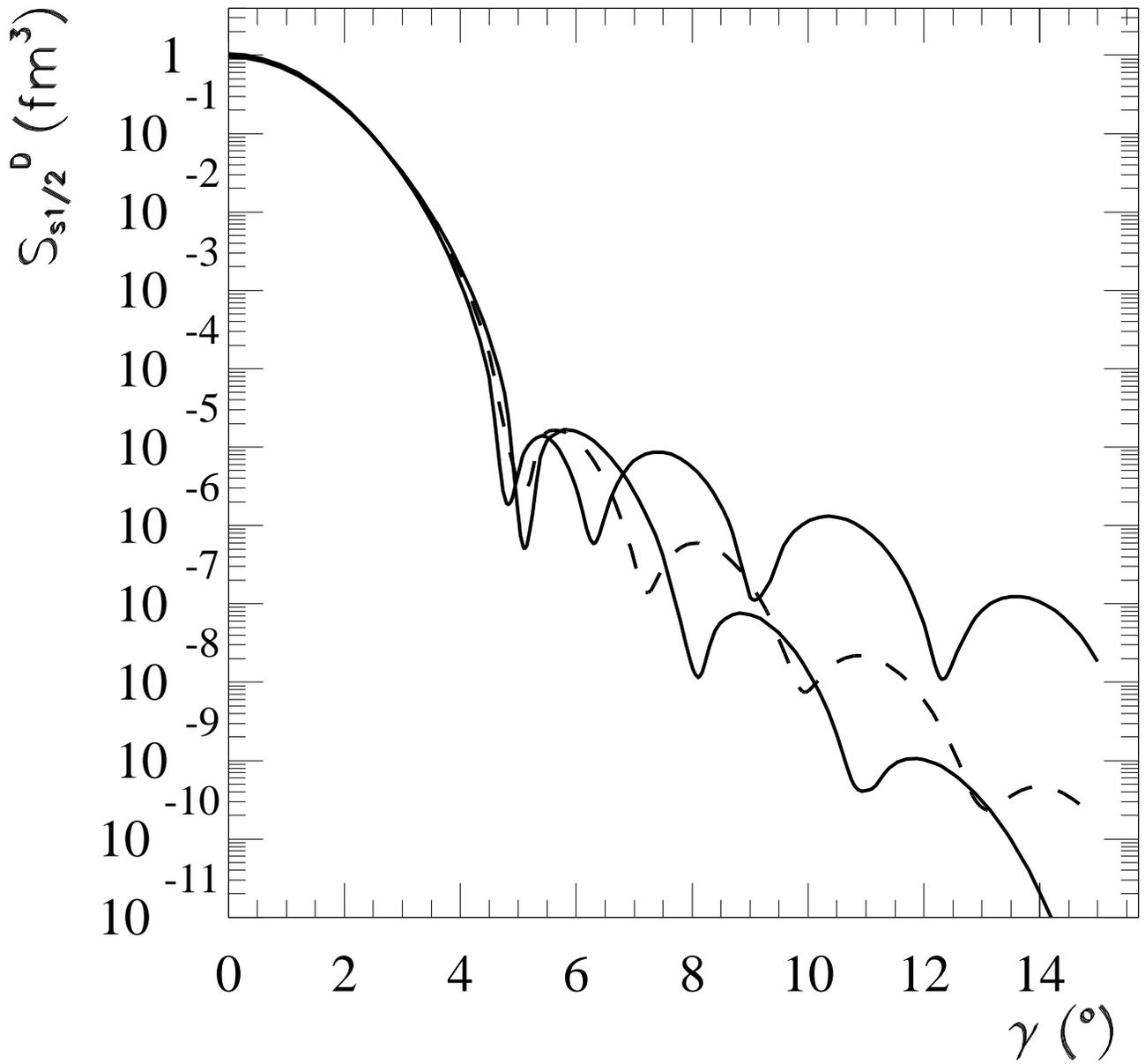
\end{document}